# Period variation of the classical Cepheid SV Vulpeculae over a century (1913-2022)

**Guy Boistel**

GEOS, Groupe Européen d'Observation Stellaire (23 parc de Levesville, 28300 Bailleau l'Evêque, France)

**Abstract**:

This study analyzes 338 new times of visual, CCD, photoelectric and photographic maxima of the classical Cepheid SV Vul. The corresponding observations were made between 1913 and 2022. On this new and large observational basis, the period variations of this star of major astrophysical interest are re-visited. Without contradicting the theory or previous studies, we show that visual observations are important for a long-term monitoring of the period variations of well-selected bright Cepheids. This study establishes the period rate of change of SV Vul at -250 s/yr. Its period is currently shorter than 45 days.

**Résumé**:

Cette étude présente 338 nouveaux instants de maxima visuels, CCD, photoélectriques et photographiques, de la céphéide classique SV Vul, collectés entre 1913 et 2022. Sur cette nouvelle et vaste base observationnelle, les variations de la période de cette étoile d'intérêt astrophysique majeur sont réexplorées. Sans bouleverser la théorie ni les études précédentes, l'étude montre que les observations visuelles sont d'une grande utilité pour un suivi à long terme des variations de période de céphéides brillantes bien sélectionnées. Cette étude établit le taux de variation de période de SV Vul à – 250 s/yr. Sa période est actuellement inférieure à 45 jours.

## 1. Introduction

It is well known that δ-Cep stars (Cepheids) define a period-luminosity relation, one of the most important extragalactical distance indicators. They also obey a fundamental relation between period and stellar age. This relation has been established thanks to the fact that some Cepheids belong to open clusters and OB associations (Hodge 1961; Efremov 1978). Thus the pulsation period of a Cepheid belonging to a cluster can be considered as an indicator of the age of the cluster if a color-magnitude diagram is not modeled.

With a period of about 45 days, the classical Cepheid SV Vulpeculae (J2000.0: 19h51m30.91s, +27°27' 36.9''), is one of them. This star belongs to the brightest Cepheids in the Milky Way (it varies between magnitudes 6.72 to 7.79 in V (GCVS, Samus et al. 2017) with one of the longest periods. This star has long been known to have the highest period variation rate, around **-230 s/yr** (Fernie 1979; Szabados 1991; Turner & Berdnikov, 2004). Finally SV Vul has been recognized as belonging to a young cluster named Alicante 13 (Turner 2014; Negueruela et al. 2020) with an estimated age of 30 My and located at a distance of 2.5 kpc. This gives SV Vul a mass of about 10 $M_\odot$ and, consequently, a luminosity close to the highest that the most recent models allow for Cepheids. It has been shown that SV Vul crosses the Cepheid instability strip for the second time (Turner & Berdnikov 2004). Table 1, taken from Turner &

**Table 1.** Representative parameters for SV Vulpeculae.

| Parameter | Value | Source |
|---|---|---|
| Period | 45$^d$0121 | Khurkarkin et al. (1985) |
| $\langle V \rangle$ | 7.209 | Berdnikov (2002) |
| $\langle B \rangle - \langle V \rangle$ | 1.462 | Berdnikov (2002) |
| Blue amplitude | 1.63 | Berdnikov (2002) |
| Reddening $E_{B-V}$ | 0.45 ± 0.01 | Turner (1984) |
| Progenitor mass | ~17 $M_\odot$ | Turner (1996) |
| Mean radius | 201.0 ± 6.0 $R_\odot$ | Turner & Burke (2002) |
| Mean $T_{eff}$ | 4830 K | Turner & Burke (2002) |
| Luminosity | $1.98 \times 10^4$ $L_\odot$ | Turner & Burke (2002) |

Berdnikov (2004), summarizes the main known features for SV Vul considered as "almost definitive" by their authors. SV Vul is thus a star of major astrophysical interest.

**2. Collecting observations**

Since the end of the years 1970s, SV Vul has been observed very regularly by GEOS members, among other classical Cepheids, which allows an examination of its long-term period behavior. These observations have not been published or only internally (Figer 1977; Dalmazzio 1997). As well as a large part of the observations made by BAV members, those of GEOS have remained largely unknown and could not be taken into account in recent studies of this star (Csörnyei et al. 2022). The availability of a large amount of numerical data now allows for a thorough study as suggested by Percy (2021) in his study on the period behavior of RU Cam.

The present study thus proposes to scan the professional classical literature and the productions of variable star observer associations to collect times of the maximum of light for SV Vul. Seven sources are used here: the old photographic and photoelectric measurements provided by the articles collected in the ADS server, and the McMaster International Cepheid Database; the CCD measurements made by Hipparcos, the KWS and ASAS-SN surveys accessible on the severs of these automatized telescopes, and also, through the AAVSO VSX server. The measurements collected by the AAVSO and accessible via the use of the VSTAR software. The times of maximum observed by the members of the German BAV group are available via their "Data for scientists" server. Finally, the observations published internally (and now in open access) by the GEOS and the systematic collection of unpublished observations since the 1980s by a few assiduous GEOS observers[1]. Our list of 338 times of maximum distributed between JD 2419903 and 2459855, provides 182 new times of maximum from visual observations made by GEOS, BAV and AAVSO, i.e., nearly 54% of the total collected times of maximum. It is quite obvious that the exploitable data for SV Vul are essentially of visual origin.

Figure 1 represents the respective contributions of the different sources of data in table 2. The full version of this table is available online at the following links:

**Table 2 - Excel .xls format.**

**Table 2 - .csv format (with commas).**

**Table 2 - .txt format (with tabulations).**

Table 2: The 338 collected times of maximum of SV Vul analyzed in this article (sample). O-C are computed on ephemeris (1).

| JJ hel 2400000+ | E(1) | O-C(1) | Quad. Res. | ADS | HIP | KWS | ASAS-SN | AAVSO | BAV | GEOS | SOURCE | Reference |
|---|---|---|---|---|---|---|---|---|---|---|---|---|
| 19903,520 | -514 | -47,15 | 0,77 | -47,15 | | | | | | | ADS | Nielsen 1937 |
| 23201,500 | -441 | -35,05 | 0,22 | -35,05 | | | | | | | ADS | Leiner 1929 |
| 23244,975 | -440 | -36,59 | -1,47 | -36,59 | | | | | | | ADS | Szabados 1991 |
| 23245,400 | -440 | -36,17 | -1,05 | -36,17 | | | | | | | ADS | Leiner 1929 |
| 23290,200 | -439 | -36,38 | -1,42 | -36,38 | | | | | | | ADS | Leiner 1929 |
| … | | | | | | | | | | | | |

Figure 2 shows the distribution between photographic (pg), Johnson V filter (V) and visual (vis.) measurements. The collected times of maximum from AAVSO is made of two sets: the 25 ones published by Berdnikov et al. (2003), and the 76 others computed by the author of this paper from the data available via the AAVSO VSTAR software.

---

[1] The internal publications of the GEOS are now in open access : http://geos.upv.es/index.php/publications/NCOA/.

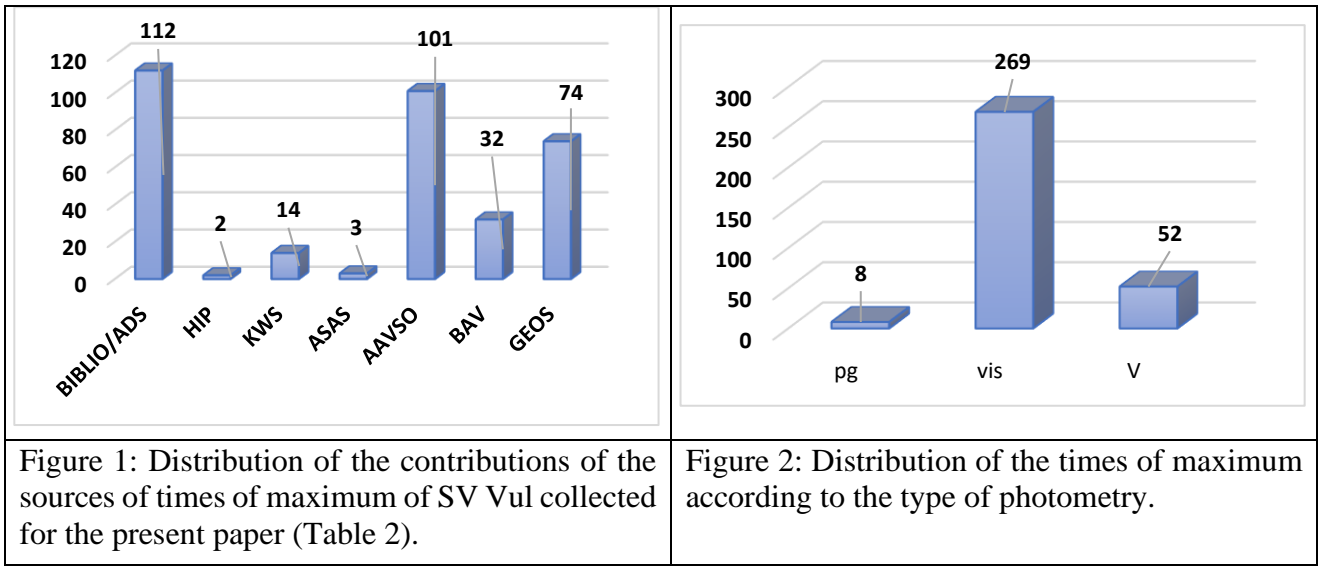

Figure 1: Distribution of the contributions of the sources of times of maximum of SV Vul collected for the present paper (Table 2).

Figure 2: Distribution of the times of maximum according to the type of photometry.

The O-C of these times of maximum have been calculated on the basis of the ephemeris given by Fernie (1979), which is currently the GCVS reference. The data concerning these times of maximum are collected in table 2: **HJD maximum = 2443086.89 + 45.0121× E (1),** E being the number of cycles elapsed since the ephemeris origin.

The data were processed with domestic Excel© procedures established, with the BAV, KWS, ASAS-SN or AAVSO-VSTAR servers returning files in .csv, .xls or .txt format. The times of maximum were primarily determined by cubic spline analysis and fitting procedures using Peranso© software.

One of the main objectives was to reproduce the shapes of the O-C diagrams published by Szabados (1991), Berdnikov (2004) and Csörnyei, Szabados et al. (2022), and then to complete them with the visual observations available at GEOS, AAVSO and BAV, data that were not used in the previously quoted studies.

A quick examination of the light curves and the available data shows that for a Cepheid type star with a period of 45 days, the average error on the determination of times of maximum is similar for visual measurements or CCD or usual Johnson V-measurements, i.e., about ± 1.5 days for its maximum value. The visual follow-up is often much more regular than the photometric monitoring as shown in the curve of figure 3 for example. In some cases, the error on the time of the maximum observed visually is about 1 day.

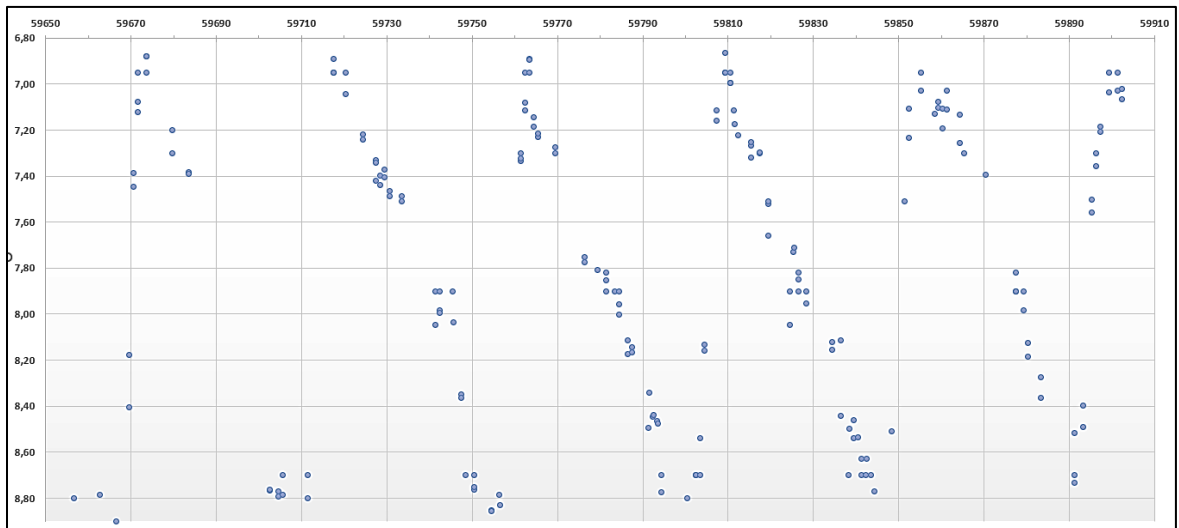

Figure 3: SV Vul, visual light curve obtained by Guy Boistel (GEOS), for the year 2022, using Nikon-Aculon 10x50 binoculars and/or Perl 12x80 binoculars (HJD +2400000).

### 3. O-C diagram and period variation of SV Vul

The time of maximum of table 2 allow to draw the diagram of O-C (in days) calculated on the ephemeris (1). Figure 4 shows the O-C of maxima the color of the points corresponding to different sources. In figure 5 the totality of the O-C is modeled by a quadratic function (the best-fit parabola). Figure 4 and 5 illustrates the very good compatibility of the visual and V maxima determinations.

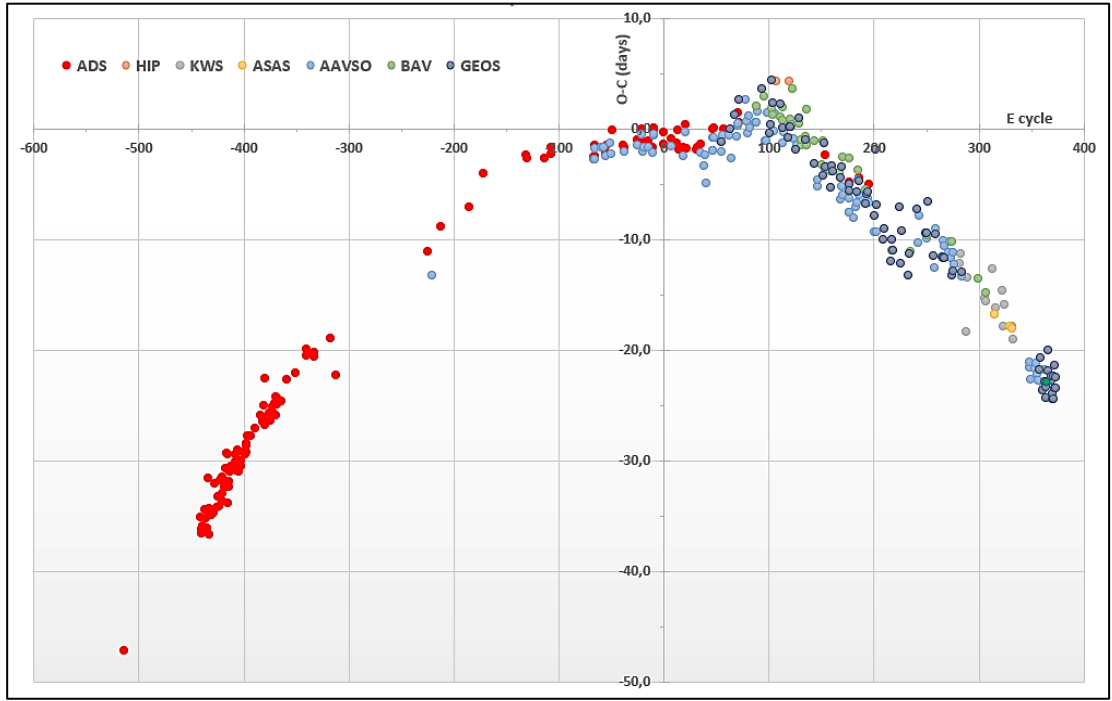

Figure 4: SV Vul, O-C diagram by source of observations, calculated on the GCVS ephemeris (1).

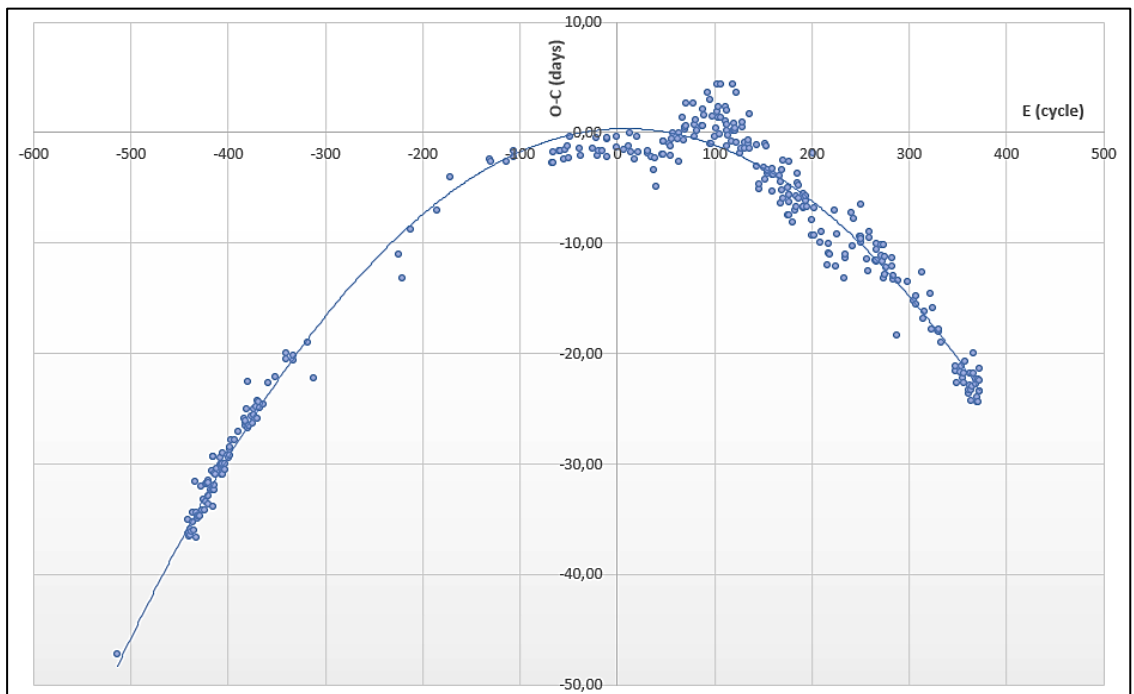

Figure 5: SV Vul, diagram of the O-C calculated on (1) and modeling by a quadratic function (2).

In figure 5 the totality of the O-C is modeled by a quadratic function (the best-fit parabola):
$$O - C\ (days) = -1.7765 \cdot 10^{-4} \times E^2 + 3.1571 \cdot 10^{-3} \times E + 0.3240\ (2).$$

Note that Csörnyei et al. (2022) has many additional points corresponding to the - 450 to - 600 cycles of our figure 5. These points are not true maximum instants but mostly "high points" extracted from the Harvard photographic plates, measurements of the magnitude close to the light maximum of SV Vul. We have preferred here to discard a large part of these "high points" in order not to disturb or even distort the study of the secular variations of the period of this Cepheid (§4). These data are available on a free server.

### 4. Long-term variations in the period of SV Vul

The modeling of figure 5 allows to obtain the quadratic term (in $E^2$) which conditions the period variation of SV Vul on the long term. It is common since the works of Parenago (1957) and Szabados (1991) in particular, to calculate the period variation ($\frac{dP}{dt}$) in sec/year in the following way (Abdel-Sabour & Sanad 2020):

$$\frac{dP}{dt}[\text{in s/yr}] = \frac{(2 \times squared\ term\ in\ E^2)}{period} \times (365.25 \times 24 \times 3600)$$

For SV Vul, with a quadratic term equal to $-1.7765 \times 10^{-4}$ (Figure 5), we obtain: - 249.13 s/yr, or about - 249 s/yr for a period of 45.0121 days. Fernie (1979) gives - 254 s/yr; Turner & Berdnikov (2004) give - 214 ± 5 s/yr in agreement with some theoretical modeling from observational data. Our value is quite consistent with previously published studies. Let us keep in mind that simply adding one time of maximum to this list causes the quadratic term to fluctuate to the order of $10^{-5}$ or $10^{-6}$ in ephemeris (2).

### 5. Irregular variations of the period of SV Vul

Figures 4 and 5 clearly show irregular variations superimposed on the mean parabola fitting the O-C. Figure 6 gives the raw curve of the quadratic residuals of the O-C on the ephemeris (2) given by the parabolic modeling in figure 5. The amplitude is well above the maximum error made on the average determination of the time of maximum (1.5 day); these fluctuations are real.

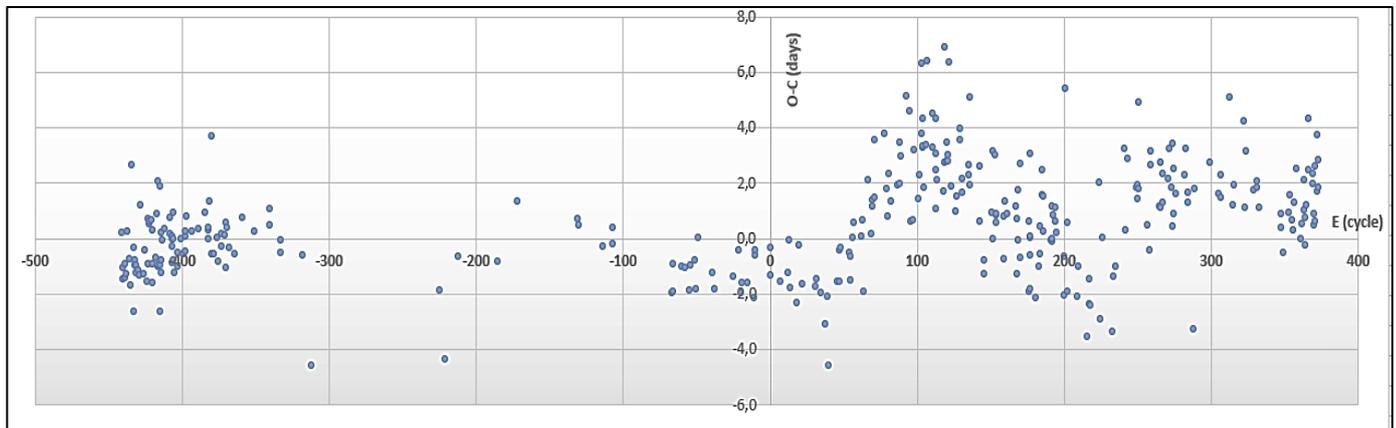

Figure 6: SV Vul, quadratic residuals of O-C in days, over the period 1913-2022.

Figure 7a gives an average curve. It is similar to the figure 7b, published by Csörnyei et al. (2022, see Figure 25 of their paper) who did not use the visual observations of GEOS and BAV (a hundred instants of maximum). Variations relative to quadratic elements up to almost 6 days are observed in the O-C quadratic residuals. These variations seem to have been more intense between the cycles +50 to +250 from the ephemeris time (1), that is to say between the Julian days 2445330 and 2454340 approximately.

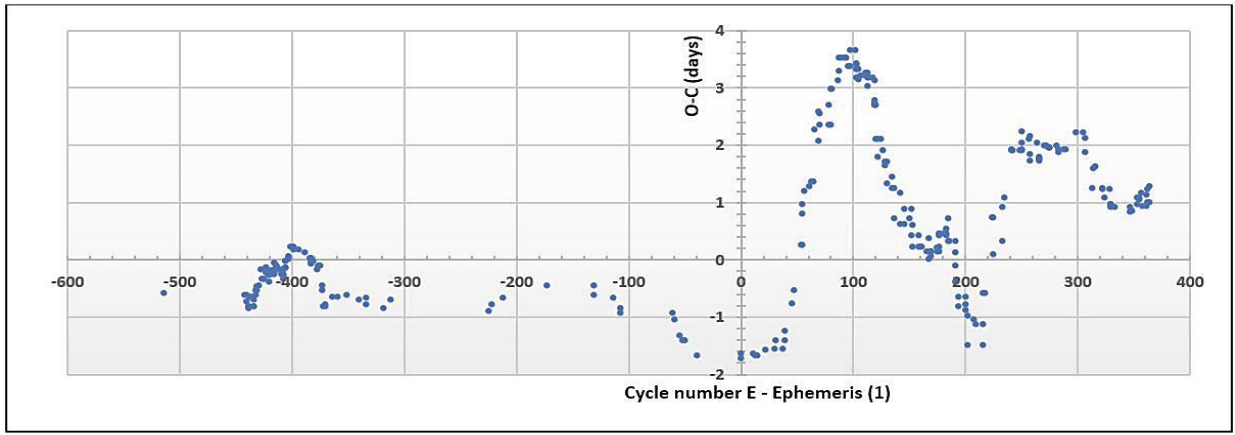

Figure 7a (This paper): SV Vul, mean square residual curve on moving averages.

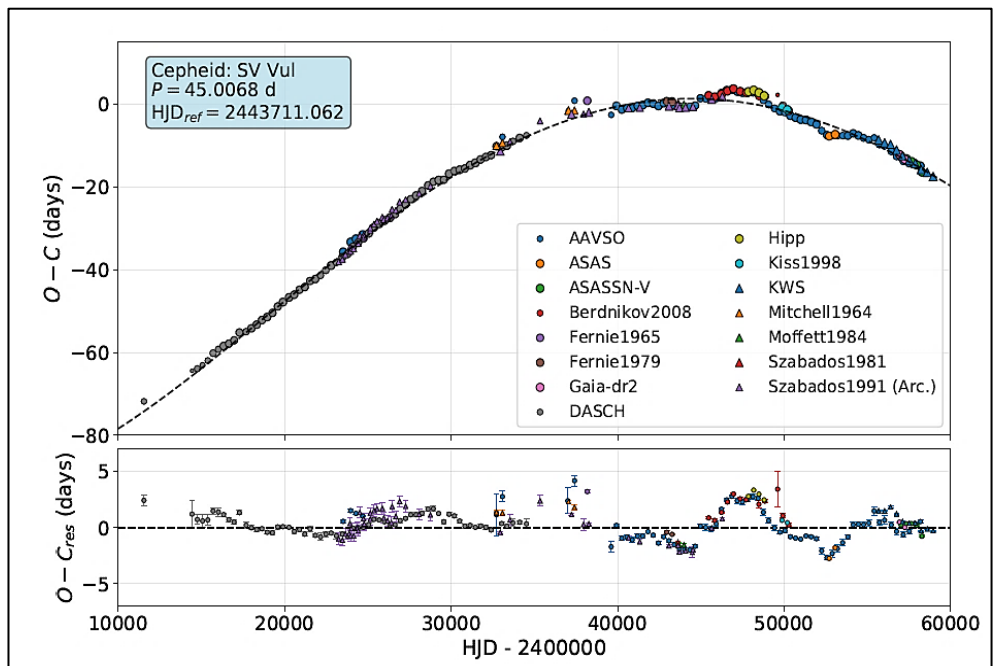

Figure 7b (Excerpt from Csörnyei et al. 2022, fig.25): O-C diagram and residual O-C diagram in HJD. The values of the phase cycles between HJD 2433000 and 2440000 (cycles between -250 and -100 of the figure 7a) have not been calculated in the same way as in this paper.

It is quite obvious that these variations are of irregular amplitude. It seems that a pseudo-periodicity can be deduced from this curve. To study this pseudo-periodicity, a period search was conducted on the average measurements used to construct figure 7, using various routines available in the Peranso© software. The observed fluctuations cannot be explained by a light-time effect (LiTE) because their amplitude is too large compared to the variation period of the star according to Szabados (1991), but also because of the irregularity of the residual O-C curve (Figure 7a). In a study developing a nonlinear hydrodynamic model in an attempt to recover the fundamental parameters of the secular fluctuations of long-period Cepheids, Fadeyev (2018) reports that the value of - 214 s/yr of the period variation $\frac{dP}{dt}$ for SV Vul, has a 10% deviation from theoretical models. But the astrophysical parameters modeled by Fadeyev (2018) ($M_{SV\ Vul} = 8.71 \times M_\odot$; $L_{SV\ Vul} = 1.09 \times 10^4 L_\odot$; $R_{SV\ Vul} = 182 \times R_\odot$) are for some of them smaller than those given by Turner & Berdnikov (2004) in table 1, and promoted by these authors as « almost definitive ».

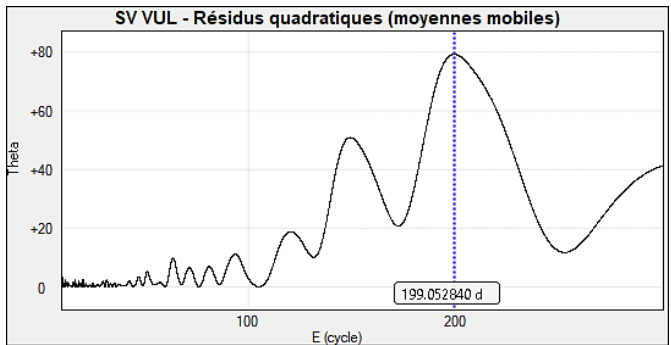 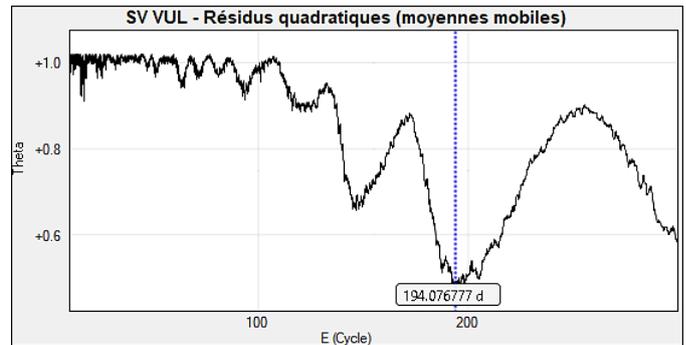

Figure 8a and 8b: SV Vul, examples of periodograms drawn on the mean squared residual variation curve (Figure 7a); Figure 8a (left) by the CLEANest method (Foster,1995); Figure 8b (right) on the PDM method (Stellingwerf, 1978) (structure 5; 2). Abscissa are given in cycles (E).

Figures 8a and 8b show samples of some periodograms we have computed with Peranso© software. All methods agree on the detection of a long-term variation of about 195-200 cycles depending on the methods used, or about 9000 days. This value is quite close to that considered by Szabados (1991). In a short note, Schröder (1978) developed an empirical third order period model with an unfortunately unjustified sinusoidal mean term, and for which he proposes a long period of about 10930 ± 581 days for the SV Vul O-C variation. However, the lack of data for the years 1930-1960 (from -300 to -50 cycles for the ephemeris (1) in figures 4 and 5), should lead us to be cautious and avoid an over-interpretation of figures 6 and 7, and of the periodograms in figures 8.

### 6. Conclusion

Without contradicting the theory, the present study of observations spread over nearly 110 years significantly completes the database of observed new times of maximum of SV Vul. In agreement with recent studies, the period variation rate calculated on a base of 338 new SV Vul times of maximum, has been specified:

$$\frac{dP}{dt} = (-249.1 \pm 1)\, s/yr.$$

The long observational time base used here provides better coverage of the irregular variations in the period of SV Vul of 5-6 days of average O-C amplitude for which a pseudo-period of about 200-205 cycles is revealed by various periodogram procedures.

We show how much can be gained from visual observations of long-period Cepheids - but not only -, and this study best answers Percy's (2021) call in his article on RU Cam: "*Skilled amateur astronomers can still make significant contributions to variable star research, even in this age of massive automated sky surveys* [...]".

SV Vul shows period variations from one cycle to another and should be observed very regularly. We recommend to visual observers to follow this star assiduously in order to determine with the highest precision the times of the maximum for a finer analysis of the small fluctuations observed and illustrated by figures 6 and 7.

**References** (this article and data of table 2):

Abdel-Sabour, M., Sanad, M., 2020, NRIAG J. Astr. Geoph., 9, 1, 99

Ahnert P., 1931, AN, 241, 265

Axelsen, R.A., 2014, JAAVSO, 42, 451

Benn, D., 2012, JAAVSO, 40, 852

Benz, W., Mayor, M., 1982, A&A, 111, 224

Berdnikov, L.N., 1986, P.Z., vol. 22, 369

Berdnikov, L.N., 1987, P.Z., vol. 22, 530

Berdnikov, L.N., Mattei, J.A., Beck, S. J., 2003, JAAVSO, 31, 146

Cragg, T.A., 1983, JAAVSO, 12, 20

Csörnyei, G., Szabados L., Molnár, L., et al., 2022, MNRAS, 511, 2125 (Data are in free access on a Wiki server)

Dalmazzio D., 1997, GEOS Notes Circulaires 854

Efremov, I. N., 1978, Soviet Astronomy, vol. 22, 161

Fadeyev, Yu. A., 2014, Astron. Lett. 40, 301

Fadeyev, Yu. A., 2018, Astron. Lett. 44, 782

Fernie, J.D., 1979, ApJ, 231, 841

Fernie, J.D., Demers, S., Marlborough, J.M., 1965, AJ, 70, 482

Figer, A., 1977, GEOS Notes Circulaires 158

Foster, G., 1995, AJ, 09, 1889

Hodge, P. W., 1961, ASPL, 8, 247

Hübscher, J., Steinbch, H-M., Vohla, F., Walter, F., 2008, OEJV, 97, 1

Kiss, L., 1998, Journal of Astronomical Data, 4, 3

Leiner, E., 1924, AN, 221, 137

Leiner, E., 1929, AN, 235, 267

Luck, R.E., Kovtyukh, V.V., Andrievsky, S.M., 2001, A&A, 373, 589

Meyer, R., 2006, OEJV, 51, 1

Mitchell, R. I., Iriarte, B., Steinmetz, D., and Johnson, H. L., 1964, Boletin de los Observatorios Tonantzintla y Tacubaya, vol. 3, 153

Nassau, J.J., Ashbrook, J., 1942, AJ, 50, 66

Negueruela, I., Dorda, R., and Marco, A., 2020, MNRAS, 494, 3028

Nielsen, A. V., 1937, AN, 262, 413

Parenago, P.P., 1956, P.Z., 11, 236

Parenago, P. P., 1957, Communications of the Konkoly Observatory, 42, 53

Paunzen, E., Vanmunster, T., 2016, AN, 337, 239

Percy, J. R., 2021, JAAVSO, 49, 1

Russo, G., Sollazzo, C., 1980, IBVS 1807

Samus, N.N., Kazarovets, E.V., Durlevich, et al., 2017, General Catalogue of Variable Stars: Version GCVS 5.1, Astronomy Reports, 61, No. 1, 80

Sanford, R. F., 1956, ApJ, 123, 201

Schröder, R., 1978, Mitteilungen der Astronomischen Gesellschaft., 43, 295

Stellingwerf, R. F., 1978, ApJ, 224, 953

Sterken, C., 2005, ASPC, 335, 3

Szabados, L., 1991, Communications of the Konkoly Observatory, 96, 123

Szabados, L., 1996, A&A, 311, 189

Turner, D.G., 1984, JRASC, 78, 229

Turner, D.G., 2014, arXiv :1403.1968v1 [astro-ph.SR]

Turner, D.G., Berdnikov, L.N., 2004,  A&A, 423, 335

Zacharov, G., 1924, AN, 221, 133

Zacharov, G., 1928, Trudy Tashkentskoj Astronomicheskoj Observatorii, 1, 33

Suran, M. D., 1990, Contributions of the Astronomical Observatory Skalnate Pleso, 20, 77